\documentclass{article}
\usepackage[utf8]{inputenc}
\usepackage{graphicx}
\usepackage{amsmath}
\usepackage{mathrsfs}
\usepackage{caption}
\usepackage{subcaption}
\usepackage[hidelinks]{hyperref}
\usepackage{enumitem}
\usepackage{multirow}
\usepackage{xcolor}
\usepackage{geometry}
\title{\textbf{An open-source alignment method for multichannel infinite-conjugate microscopes using a ray transfer matrix analysis model}}

\author{\textbf{Gemma S. Cairns\textsuperscript{1} and Brian R. Patton\textsuperscript{1*}}\\
\small 1. Department of Physics and SUPA, University of Strathclyde, Glasgow G4 0NG, UK\\
\small *brian.patton@strath.ac.uk
\date{\today}}

\begin{document}

\maketitle

\section*{Abstract}

Multichannel, infinite-conjugate optical systems easily allow implementation of multiple image paths and imaging modes into a single microscope. Traditional optical alignment methods which rely on additional hardware are not always simple to implement, particularly in compact open-source microscope designs. We present here an alignment algorithm and process to position the lenses and cameras in a microscope using only image magnification measurements. We show that the resulting positioning accuracy is comparable to the axial resolution of the microscope. Ray transfer matrix analysis is used to model the imaging paths when the optics are both correctly and incorrectly aligned. This is used to derive the corresponding image magnifications. We can then extract information about the lens positions using simple image-based measurements to determine whether there is misalignment of the objective lens to sample distance (working distance) and with what magnitude and direction the objective lens needs to be adjusted. Using the M4All open-source 3D printable microscope system in combination with the OpenFlexure microscope, we validate the alignment method and highlight its usability. We provide the model and an example implementation of the algorithm as an open-source Jupyter Notebook.

\section{Introduction}

\begin{figure}[h!]
    \centering
    \includegraphics[width=9cm]{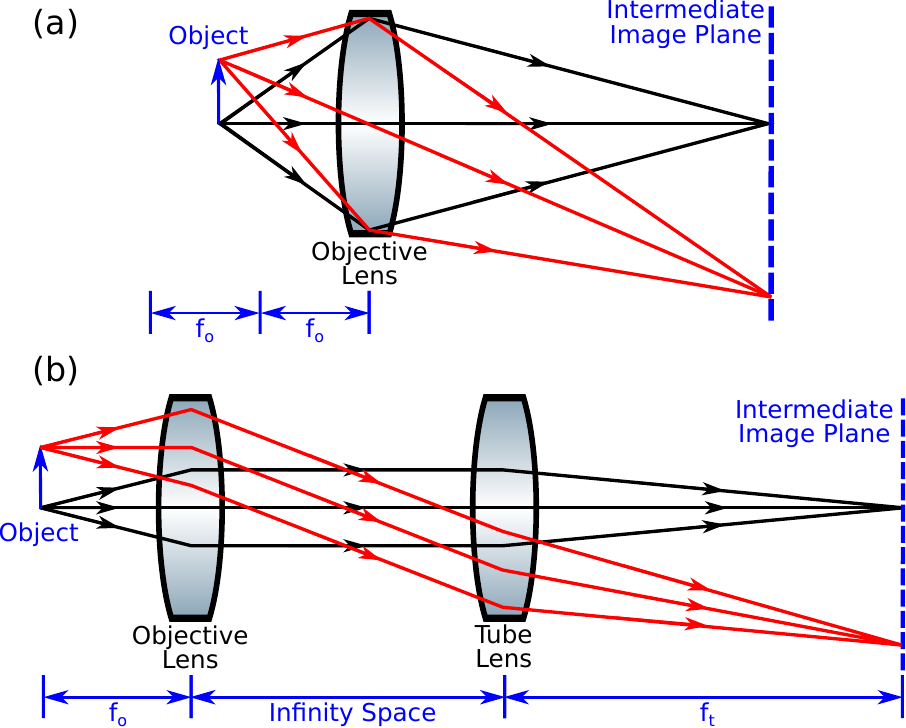}
    \caption{Optical schematics of (a) a finite-conjugate optical system where the sample is positioned between $f_o$ and $2f_o$ and the objective lens focuses and forms an image at a finite distance, and (b) an infinite-conjugate optical system where the sample is positioned at $f_o$ and objective lens does not directly form a real image of the object, therefore a tube lens with focal length $f_t$ must be used to focus the collimated beam. The collimated path is referred to as the infinity space. Figure from \cite{cairns_development_2022}.}
    \label{FinitevsInfinite}
\end{figure}

Advanced microscopes often include multiple optical paths in the system to enable, for example, multi-colour fluorescence microscopy and to combine multiple modes of microscopy in one instrument. There are two different ways to design an imaging path; finite-conjugate and infinite-conjugate systems. In a finite-conjugate design the sample is positioned between $f_o$ and 2$f_o$ before the objective lens (where $f_o$ is the effective focal length of the objective lens). The objective lens then focuses light at an intermediate image plane \cite{sanderson_basic_2019} where either an imaging sensor or relay lens, such as an eyepiece for direct observation, can be placed (figure \ref{FinitevsInfinite} (a)). While there are methods for allowing multiple imaging paths in a finite-conjugate system, it can be more complicated to implement than in an infinite-conjugate system.

An infinite-conjugate system positions the sample at $f_o$ before the objective lens and produces a collimated beam after the objective lens from a single point source in the focal plane. The collimated beam subsequently must be focused using a tube lens to form an image (figure \ref{FinitevsInfinite} (b)) \cite{sanderson_basic_2019}. To easily implement multiple imaging paths, non-focusing optics for splitting light into different detection channels, such as dichroic mirrors and beamsplitters, can be placed between the objective and tube lenses in the collimated ``infinity space'' without introducing spherical aberration into the system and without changing the position of the image plane \cite{sanderson_microscope_2019}. The distance between the objective and tube lens can also be varied without impacting the magnification, further easing the implementation of multichannel systems. 

\begin{figure}[h!]
    \centering
    \includegraphics[width=5cm]{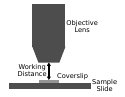}
    \caption{Illustration of the definition of the working distance of an objective lens with reference to the sample coverslip. Note that, for objectives designed to be used with a coverslip, the imaging plane in this example will be immediately under the coverslip. Figure from \cite{cairns_development_2022}.}
    \label{fig:WorkingDistance}
\end{figure}

Note that, figure \ref{FinitevsInfinite} depicts the objective lenses as single lens elements, however in practice objective lenses contain multiple lens elements. Therefore, $f_o$ is measured from an effective plane within the objective lens body, which is not normally marked. Instead, objective lens manufacturers also state a working distance for the lens, which is  illustrated in figure \ref{fig:WorkingDistance}. For objectives designed to work with a coverslip, the working distance does not include the coverslip thickness (i.e. an objective specified to have a 400\,$\mu$m working distance and working with 170\,$\mu$m coverslips will have the focal plane 570\,$\mu$m from the front surface of the objective). For an objective placed at the working distance from the coverslip, as in figure \ref{fig:WorkingDistance}, this means that the focal plane will be coincident with the bottom surface of the coverslip.

When setting up a multichannel infinite-conjugate microscope, it is important that the objective lens - tube lens system is aligned so that the plane being imaged on the sensor is positioned at the correct working distance. If it is not, the infinity space will not be collimated, resulting in different magnifications for each channel if they have different path lengths. In addition, other aberrations and field distortions may be introduced when using the objective at the wrong working distance.

Note that collimation refers to light emitted from a point source. As can be seen in figure \ref{FinitevsInfinite}\,(b), extended objects in the sample plane will result in beam divergence in infinity space, as each collimated bundle of rays from each point contributing to the extended object will propagate at a different angle to the optical axis (compare the black and red ray bundles). Therefore, for microscopes that image samples with illumination spread over a wide area, collimation is not the same as looking to see if all the light rays coming out of the back of the objective remain parallel. Instead, checking whether the rays from a point source are collimated (i.e. checking that the sample is at the correct working distance in an infinite-conjugate system) must be achieved through an appropriate technique. Traditional methods, such as using an auto-collimator \cite{johnstone_using_2022} or shear plate, are very effective, but require dedicated hardware. Recently there has been a growing community of researchers focused on developing open-source hardware for microscopy (see \cite{noauthor_hohlbeinlabopenmicroscopyom_hardware_2022} for an extensive list of projects), where designs are becoming increasingly compact which results in difficulties using traditional optical alignment methods. For example, an auto-collimator may not fit into the optical path. Therefore, to align an infinite-conjugate multichannel microscope without the use of additional hardware, we have developed an image-based alignment method based on a mathematical ray transfer matrix analysis model. The only requirements for the method are:
\begin{itemize}
    \item To have a way of accurately controlling the z step (focusing) movements of either the sample or objective lens.
    \item To know the specifications of the optics and cameras in the system.
    \item To decide whether it is important to know the absolute magnifications of the imaging channels or whether it is adequate to know their relative magnifications to one another. This will determine whether a feature-size calibrated sample is required, e.g. a calibrated graticule slide.
\end{itemize}
    
In this paper we describe the mathematical model and resulting alignment method in detail before showing it being used to align a low-cost, open-source and 3D printable multichannel microscope built using M4All \cite{cairns_nanobiophotonics-strathclydem4all_2021} combined with the OpenFlexure microscope stage \cite{sharkey_one-piece_2016}. 

\section{Ray Transfer Matrix Analysis Theory}

Mathematical ray tracing calculations within the paraxial approximation (where only light rays which make a small angle to the optical axis are considered, such that $sin(\theta)\approx\theta$) can be performed using ray transfer matrix analysis (RTMA). Note that in high numerical aperture ($N\!A=nsin(\theta)$) systems, where $\theta$ is larger, ray-tracing still often gives useful results. With this in mind, we can recommend this alignment approach even with high-$N\!A$ ($N\!A>$\,0.6) objectives. For those unaware of the theory of RTMA, reference\,\cite{valerie_pineau_noel_tools_2021} provides an excellent introduction to both the theory and the Python library we use in this paper.

A light ray at a plane along the optical axis, $z$, has a height, $y$, and angle $\theta$, with respect to $z$ which is represented as a ray vector:

\begin{equation}
    \boldsymbol{r} =
    \begin{bmatrix}
    y \\
    \theta
    \end{bmatrix}
\end{equation}

In RTMA the input ray vector is transformed through different optical elements or free space propagation paths which are described by 2x2 matrices, known as transfer matrices and also often referred to by their indices as ABCD matrices. The output ray vector is defined by left multiplication of the input ray vector with the transfer matrices for each element (note here that the ABCD matrix represents the transfer matrix for the total system):

\begin{equation} \label{ABCD Analysis}
    \begin{bmatrix}
        y_{out} \\
        \theta_{out}
    \end{bmatrix}
    =
    \begin{bmatrix}
        A & B \\
        C & D
    \end{bmatrix}
    \begin{bmatrix}
        y_{in} \\
        \theta_{in}
    \end{bmatrix}
    =
    \begin{bmatrix}
        Ay_{in}+B\theta_{in} \\
        Cy_{in}+D\theta_{in}
    \end{bmatrix}
\end{equation}

For this work it is sufficient to use only the transfer matrices for free space and a thin lens respectively, where $d$ is the propagation distance in free space and $f$ is the focal length of the thin lens (transfer matrices for further elements and matrix derivations can be found in Burch \textit{et al.} \cite{gerrard_introduction_1994}):

\begin{equation} \label{ABCDfreespace}
    \begin{bmatrix}
        1 & d \\
        0 & 1
    \end{bmatrix}
\end{equation}

\begin{equation} \label{ABCDlens}
    \begin{bmatrix}
        1 & 0 \\
        -\frac{1}{f} & 1
    \end{bmatrix}
\end{equation}

The total ABCD matrix can be used to derive some useful properties of the system \cite{valerie_pineau_noel_tools_2021,gerrard_introduction_1994}. Most importantly for this work is the fact that when $B=0$ the system produces a real image at the output plane from an object at the input plane. This is equivalent to $y_{out}$ being independent of $\theta_{in}$. The lateral and angular magnifications in this case are given by $A$ and $D$ respectively.

\section{Ray Transfer Matrix Analysis Model for Alignment of Infinite-Conjugate Microscope Designs}

\begin{figure}[h!]
    \centering
    \includegraphics[width=7.5cm]{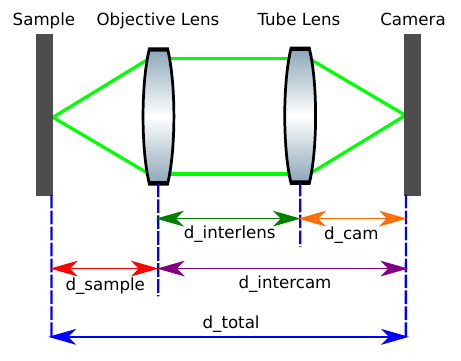}
    \caption{Simplified schematic of a single infinite-conjugate imaging channel to define variables for ray transfer matrix analysis. Figure from \cite{cairns_development_2022}.}
    \label{fig:SingleChannelDiagram}
\end{figure}

The systems we wished to align comprised of a single objective and an additional single tube lens per optical path, as shown in figure \ref{fig:SingleChannelDiagram}. We therefore demonstrate the application of our alignment routine for such a microscope - we anticipate that it would also work for more complex optical paths with suitable calculation of the total ABCD matrix. To model a correctly aligned microscope with an infinite-conjugate optical design, we define variables in figure \ref{fig:SingleChannelDiagram}. The total length of the channel from the sample to the camera sensor is $d\_total$, and due to the design of the OpenFlexure microscope and the M4All system, will be treated as being fixed in the following for this example system. For other systems, the total distance may change with sample positioning and this would need to be incorporated in the calculation of the total ABCD matrix. Treating the objective lens as a single thin lens, the distance between the sample and the objective lens is $d\_sample$. The distance between the objective lens and the camera sensor, and the objective lens and the tube lens is $d\_intercam$ and $d\_interlens$ respectively. Finally, we define the distance between the tube lens and the camera sensor as $d\_cam$. This set of variables implies the manner in which we align the system - the sample is placed as close to the correct working distance as we can estimate by moving the objective (the sample stage is fixed in position with respect to the propagation axis), the camera is also fixed in position at $d\_total$ from the sample by a non-adjusting mount, and we move the tube lens to focus the image of the sample.

The defined variables, along with the objective lens effective focal length $f_o$ and tube lens focal length $f_t$, can be substituted into the ABCD matrices \ref{ABCDfreespace} and \ref{ABCDlens} to build the matrix equation \ref{ABCD Infinity System} for the infinite-conjugate imaging channel (where the matrices are left multiplied in the order they are positioned in the optical path).

\begin{equation} \label{ABCD Infinity System}
    \begin{bmatrix}
        y_{out} \\
        \theta_{out}
    \end{bmatrix}
    =
    \begin{bmatrix}
        1 & d\_cam \\
        0 & 1
    \end{bmatrix}
    \begin{bmatrix}
        1 & 0 \\
        -\frac{1}{f_t} & 1
    \end{bmatrix}
    \begin{bmatrix}
        1 & d\_interlens \\
        0 & 1
    \end{bmatrix}
    \begin{bmatrix}
        1 & 0 \\
        -\frac{1}{f_o} & 1
    \end{bmatrix}
    \begin{bmatrix}
        1 & d\_sample \\
        0 & 1
    \end{bmatrix}
    \begin{bmatrix}
        y_{in} \\
        \theta_{in}
    \end{bmatrix}
\end{equation}
\vspace{0.25cm}

The following definition can also be made for $d\_interlens$:

\begin{equation} \label{interlens}
    d\_interlens = d\_total - d\_sample - d\_cam
\end{equation}

In a correctly aligned system, $d\_sample$ is equal to $f_o$ and $d\_cam$ is equal to $f_t$. Therefore the matrix equation becomes:

\begin{equation} \label{ABCD Infinity System2}
    \begin{bmatrix}
        y_{out} \\
        \theta_{out}
    \end{bmatrix}
    =
    \begin{bmatrix}
        1 & f_t \\
        0 & 1
    \end{bmatrix}
    \begin{bmatrix}
        1 & 0 \\
        -\frac{1}{f_t} & 1
    \end{bmatrix}
    \begin{bmatrix}
        1 & d\_total - f_o - f_t \\
        0 & 1
    \end{bmatrix}
    \begin{bmatrix}
        1 & 0 \\
        -\frac{1}{f_o} & 1
    \end{bmatrix}
    \begin{bmatrix}
        1 & f_o \\
        0 & 1
    \end{bmatrix}
    \begin{bmatrix}
        y_{in} \\
        \theta_{in}
    \end{bmatrix}
\end{equation}
\vspace{0.25cm}

Upon substituting the microscope design values for $d\_total$, $f_o$ and $f_t$ into the matrix equation for a correctly aligned system and multiplying the transfer matrices to obtain a single transfer matrix for the total channel, the lateral magnification of the image, $A$, will equal the value $M$ obtained using:

\begin{equation} \label{Eq:Magnification2}
    M = \frac{f_t}{f_o}
\end{equation}

However, the magnification of an incorrectly aligned microscope, such as when the objective lens is not at the correct working distance, will differ from equation \ref{Eq:Magnification2}. This is because when $d\_sample \neq f_o$ an image can only be formed when $d\_cam \neq f_t$. The first step in our calibration routine is therefore to calculate the magnification for each channel for a range of suitable $d\_sample$ values, centred around the real effective focal length of the objective lens, $f_o$. To do this calculation we substitute equation \ref{interlens} into equation \ref{ABCD Infinity System}, set a value for $d\_sample$ from the range chosen as appropriate for the objective, and solve for the value of  $d\_cam$ that gives an image at the sensor. This is easily done by recalling that the $B$ component of the ABCD matrix of a system is equal to zero for systems producing a real image at the output plane from an object at the input plane. Therefore a function that returns the value of B for a given physical setup can be passed to e.g. the Python fsolve routine allowing a numerical solution for the value of $d\_cam$ that produces an image for each $d\_sample$ in the range of interest. Note that it is possible that no imaging solution can be found, given the fixed camera position and the choice of $d\_sample$ range. In this case, our example code fails gracefully and warns the user of the position at which the failure occurs for the relevant path. 

Substituting the solved $d\_cam$ value for each $d\_sample$ value back into the matrix equation allows the lateral magnification $A$ to be determined for each iteration. A plot of lateral magnification vs $d\_sample$ shows the deviation in magnification when the objective lens is moved away from the correct working distance. For a multichannel microscope, repeating the analysis for each channel allows the theoretical difference in magnification between each channel to be modelled in the situation where the objective lens is not positioned correctly along the optical path. Note that the fundamental design choice that creates the differing magnification is the different path lengths for each channel. As such, if the microscope is designed with equal path lengths, it may be worthwhile to introduce a temporary path difference to allow alignment using this method. Since both arms will then be set up after alignment to image at the correct working distance, the path difference can be subsequently removed from the modified arm and that arm corrected relative to the unmodified arm.

\begin{figure}[h!]
    \centering
    \includegraphics[width=10.5cm]{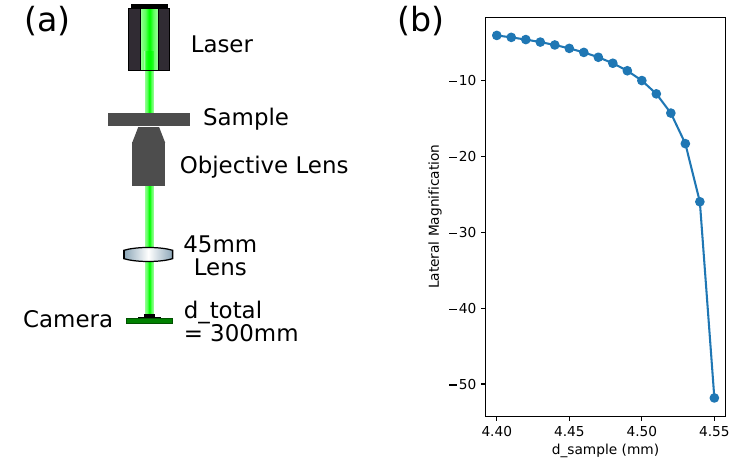}
    \caption{(a) Schematic of a single channel infinite-conjugate microscope. Figure from \cite{cairns_development_2022}. (b) Plot of lateral magnification vs $d\_sample$ created using our ray transfer matrix model for an infinite-conjugate microscope, setting $d\_total$ to 400\,mm, $f_o$ to 4.5\,mm and $f_t$ to 45\,mm.}
    \label{fig:SingleChannelplot}
\end{figure}

The theoretical plot of lateral magnification vs $d\_sample$ can be used to align the position of the objective lens and tube lenses in an infinite-conjugate microscope. For a single channel microscope the practical magnification of the microscope can be measured using for example a graticule sample. Then using this value, $d\_sample$ can be interpolated from the theoretical plot (example plot shown in figure \ref{fig:SingleChannelplot}). The difference in distance between $d\_sample$ and $f_o$ is the distance the objective lens needs moved to be at the correct working distance and thus correcting the magnification of the microscope to the expected value.

\begin{figure}[h!]
    \centering
    \includegraphics[width=13cm]{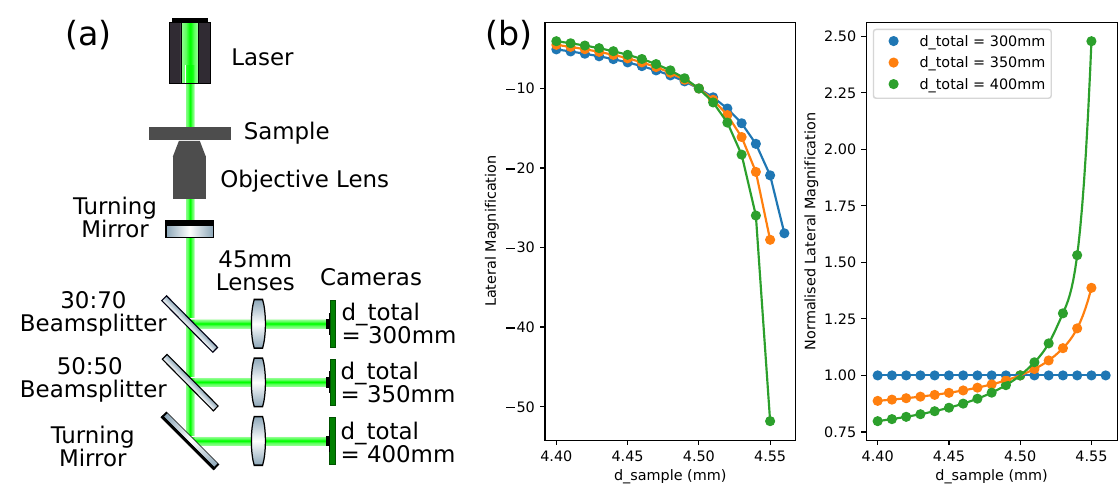}
    \caption{(a) Schematic of three channel infinite-conjugate microscope. Figure from \cite{cairns_development_2022}. (b) Plots of lateral magnification and normalised lateral magnification vs $d\_sample$ created using our ray transfer matrix model for a three channel infinite-conjugate microscope, setting  $d\_total$ to 300\,mm, 350\,mm and 400\,mm, $f_o$ to 4.5\,mm and $f_t$ to 45\,mm for each channel.}
    \label{fig:MultiChannelplot}
\end{figure}

We focus here, however, on alignment of multichannel infinite-conjugate microscopes in the case where a calibration sample, such as a graticule, may not be available. In this case, a plot of the modelled lateral magnification vs $d\_sample$ for each channel is created and then, so as to mitigate the need to measure true magnifications, each plot is normalised to a single channel (we select the channel with the shortest path length for consistency) to create a plot of normalised magnification vs $d\_sample$ for each channel. An example of the plots is given in figure \ref{fig:MultiChannelplot} for a multichannel microscope with three channels where the path lengths of each channel are $d\_total$ = 300\, mm, 350\,mm and 400\,mm respectively. All three channels are otherwise identical in terms of cameras and tube lenses. The camera specification of note is the pixel size; we measure the size of identifiable features within the image, and from there estimate the relative magnification for each channel, using image pixel distances. Therefore, different sized camera pixels will be measuring different absolute distances on the imaging plane and must be compensated for when comparing images from different pixel-sized cameras in the same microscope. 

Note here that it can be seen that at the point where $d\_sample$ = $f_o$ = 4.5\,mm, the three channels have equal magnifications as expected. Then, as the objective lens is moved away from the correct position, the magnifications vary from one another. When the objective lens is too close to the sample ($d\_sample <$\,4.5\,mm) the third channel has the smallest magnification. Whereas the first channel has the smallest magnification when the objective lens is too far away from the sample ($d\_sample >$\,4.5\,mm). This allows unambiguous estimation of both the magnitude and direction of a positioning error of the sample plane.

To use the calculated plots for a multichannel infinite-conjugate microscope to align the objective lens and tube lenses the following steps are followed:

\begin{enumerate}
    \item Set up the microscope with each channel in focus.
    \item Capture an image on each channel of a sample where the distance in pixels between the same two points can be measured (a specific calibration slide allows absolute magnifications to be calibrated and measured, while a general sample will allow for alignment but not confirm the final magnification).
    \item Normalise the distances measured on the calibration images to the distance measured on the channel which was used for the lateral magnification normalisation calculations. 
    \item To determine the predicted error in the position of the objective lens, the measured normalised magnification value for the channel with the largest $d\_total$ value, can be plotted on the normalised magnification graph and the corresponding $d\_sample$ value can be interpolated, which we define as $d\_sample\_interpolated$. The error in the position of the objective lens is then calculated according to equation \ref{workingdistanceerror}. If the working distance error is a positive value then the objective lens is that magnitude too far away from the sample, and if it is a negative value then it is that magnitude too close to the sample.
\end{enumerate}

\vspace{-0.5cm}
\begin{equation} \label{workingdistanceerror}
    working\ distance\ error = d\_sample\_interpolated - f_o 
\end{equation}

There are some considerations to be made with this approach. Since we are normalising distances relative to an ideal system, we are making some important assumptions e.g. all lenses have the design focal length with no consideration of manufacturing tolerances. As such, there are a few areas where it's worth applying a critical approach when working with this alignment algorithm. Large discrepancies in the estimated current value of $d\_sample$ between different channels could indicate some of the following issues:

\begin{itemize}
    \item The measurement of the feature size within the image implicitly assumes an image with no distortions. To minimise the impact of any distortions that are present, try to get feature size measurement from a region central to the image in case the magnification differs over the field of view (typical with high magnification, very simple optical systems). This is the easiest problem to test for, since it just requires repeating the computational side of the alignment, without needing new images.
    \item Tolerance differences on tube lenses (a 45\,mm nominal focal length lens might have a different focal length as manufactured) are the final source of error we consider. This is likely the source of absolute errors on calibration (when all paths agree on the magnification, but it differs from the expected magnification, or when the calculated $d\_sample$ position is significantly different on each path). It is slightly more likely to be observed when a range of different tube lenses are used (e.g. same focal lengths but different lens types or different focal lengths to suit different cameras). This is the hardest to ascertain on a purely image-based system of calibration - it may be that testing of the focal length is required for each lens.
\end{itemize}

Finally, we note that the use of normalised magnification also allows the alignment of channels that have different absolute magnifications and/or cameras with differing pixel sizes. If the normalisation is performed over both relative magnification and relative pixel size, then the error in $d\_sample$ can still be estimated. See our sample code for an example of how to implement this normalisation.

\section{Practical Examples}

\begin{figure}[h!]
    \centering
    \includegraphics[width=12cm]{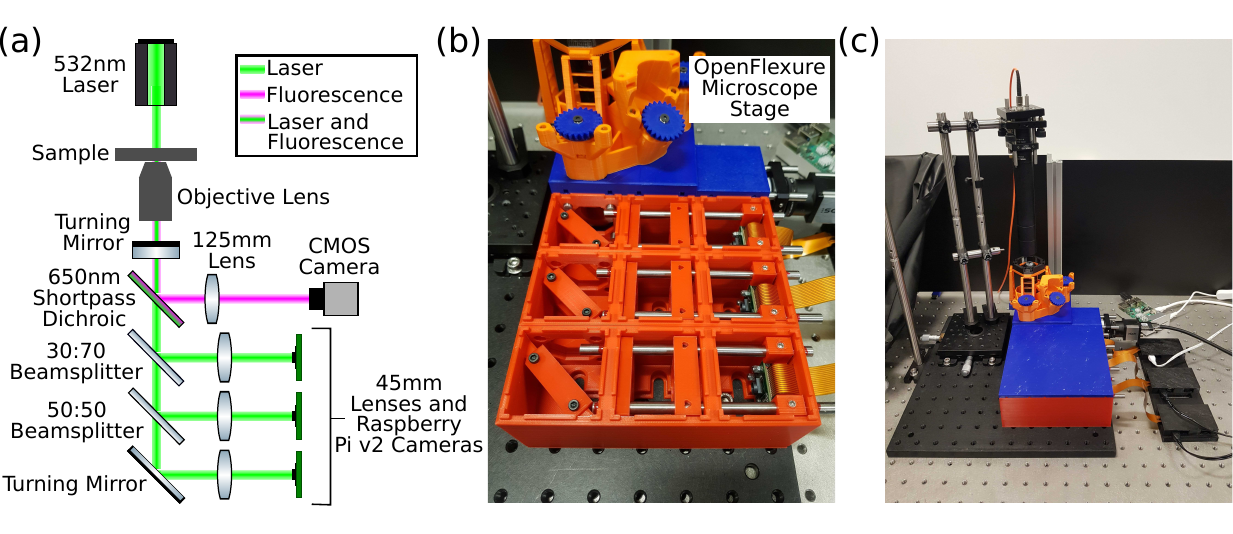}
    \caption{(a) Schematic of the M4All Fluorescence and TIE Microscope where the three brightfield channels have path lengths of $d\_total$ = 300, 350 and 400\,mm respectively. (b) Zoomed in photo of the M4All microscope in practice in combination with the OpenFlexure microscope stage \cite{sharkey_one-piece_2016}. (c) Zoomed out photo of the overall microscope including the illumination optics and the Raspberry Pi controllers. Note that the three axes on the OpenFlexure stage were controlled using stepper motors but are not shown here. Figure adapted from \cite{cairns_development_2022}.}
    \label{fig:TIEandFluorescenceMicroscope}
\end{figure}

To show the alignment procedure works in practice, we used the M4All Fluorescence and TIE Microscope, figure \ref{fig:TIEandFluorescenceMicroscope}. M4All is an open-source 3D printable microscope system \cite{cairns_nanobiophotonics-strathclydem4all_2021} which is compatible with the OpenFlexure microscope \cite{sharkey_one-piece_2016}. Full build instructions can be found on the M4All repository \cite{cairns_nanobiophotonics-strathclydem4all_2021}. Briefly, this microscope was designed for single channel fluorescence and simultaneous brightfield multifocal plane imaging to enable computational phase contrast microscopy using the transport of intensity equation (TIE) \cite{davis_transport_2017}. A schematic of the microscope can be seen in figure \ref{fig:TIEandFluorescenceMicroscope} (a) along with photos of the microscope in figures \ref{fig:TIEandFluorescenceMicroscope} (b) and (c).

A turning mirror (Thorlabs PF10-03-P01) placed below the OpenFlexure microscope stage couples the light into the M4All cubes. A 650\,nm shortpass dichroic mirror (Thorlabs DMSP650) then reflects fluorescence emission $\geq$ 650\,nm which is focused by a 125\,mm focal length tube lens (Thorlabs AC254-125-A) onto an IDS CMOS camera (UI-3060CP-M-GL Rev. 2). The remaining transmitted laser light is split into the three brightfield channels by 30:70 and 50:50 beamsplitters (Thorlabs BSS10R and BSW10R) and focused by 45\,mm focal length tube lenses (Thorlabs AC254-045-A) onto Raspberry Pi v2 camera modules. The three brightfield channels have the same $d\_total$ values of 300, 350 and 400\,mm as the example in figure \ref{fig:MultiChannelplot}. Before altering the positions of the three brightfield channel tube lenses to enable multifocal plane imaging for future work, we first used the ray transfer matrix alignment method to co-align each channel to image at the correct working distance. 

\begin{figure}[h!]
    \centering
    \includegraphics[width=11cm]{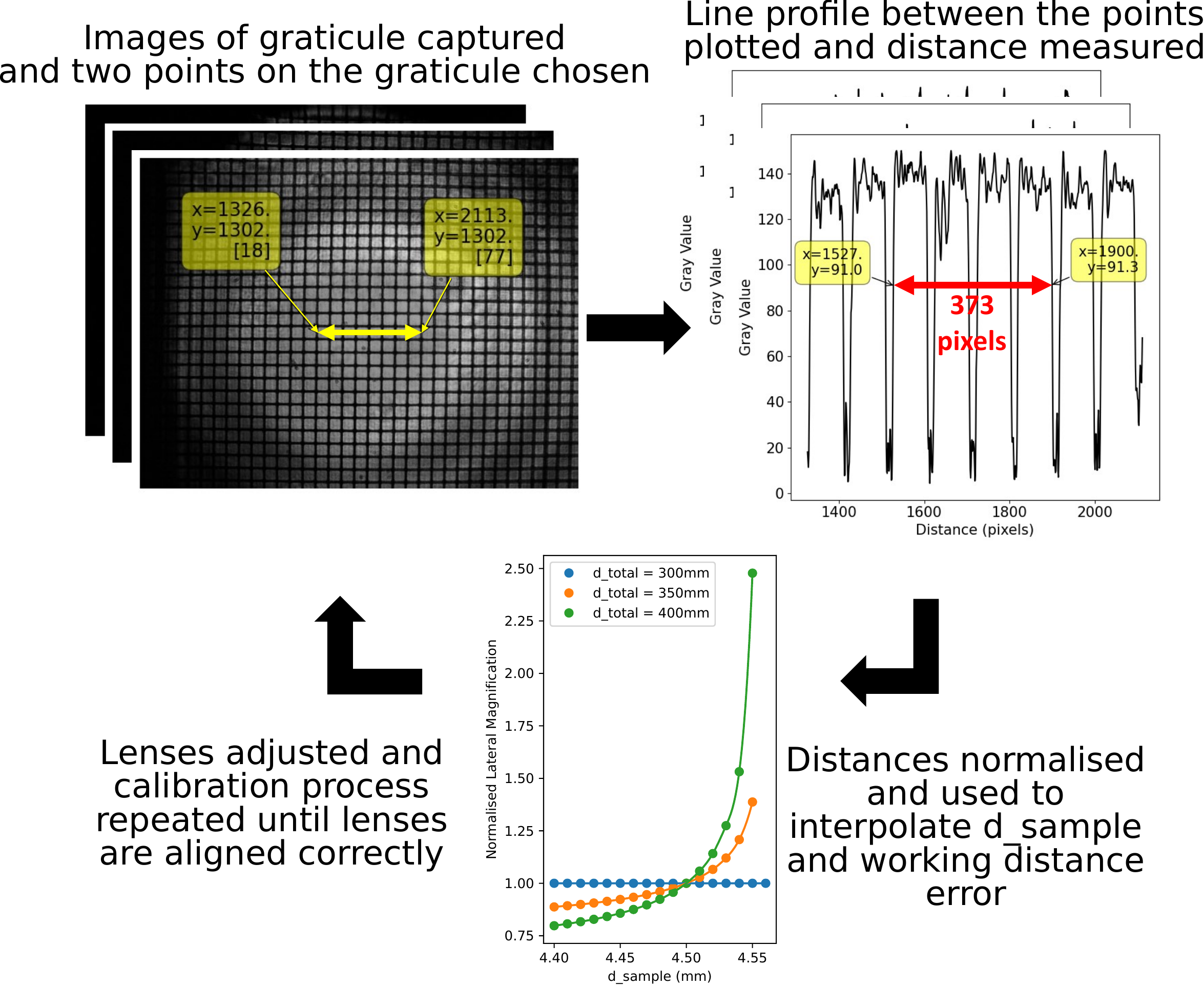}
    \caption{Outline of alignment process steps where an image of a 10\,$\mu$m graticule is captured on each channel. Next a line profile is plotted for each image and the distance between the same number of grid squares is measured in pixels. The distances are then normalised and the plot of lateral magnification vs $d\_sample$ obtained from the ray transfer matrix analysis model is used to interpolate $d\_sample$ and thus the objective lens' working distance error. Finally the lenses are re-positioned and the alignment process repeated until the lenses are aligned. Figure adapted from \cite{cairns_development_2022}.}
    \label{fig:calibrationprocess}
\end{figure}

We first set the microscope up with each brightfield channel in focus and captured an image of a 10\,$\mu$m graticule sample (as discussed however, any sample where the distance in pixels between the same two feature points can be measured for every channel can be used). As we were using a graticule, a line profile of the graticule was plotted for each channel and the distance in pixels between the same two points on each plot was measured. The distances were then normalised to the first brightfield channel ($d\_total$ = 300\,mm), which we call the measured normalised magnifications, and used to interpolate $d\_sample$ from the plot of normalised lateral magnification vs $d\_sample$ in figure \ref{fig:MultiChannelplot} (b). The working distance error was then calculated from equation \ref{workingdistanceerror}. A flow chart of the alignment steps is given in figure \ref{fig:calibrationprocess}.

\begin{figure}[h!]
    \centering
    \includegraphics[width=13cm]{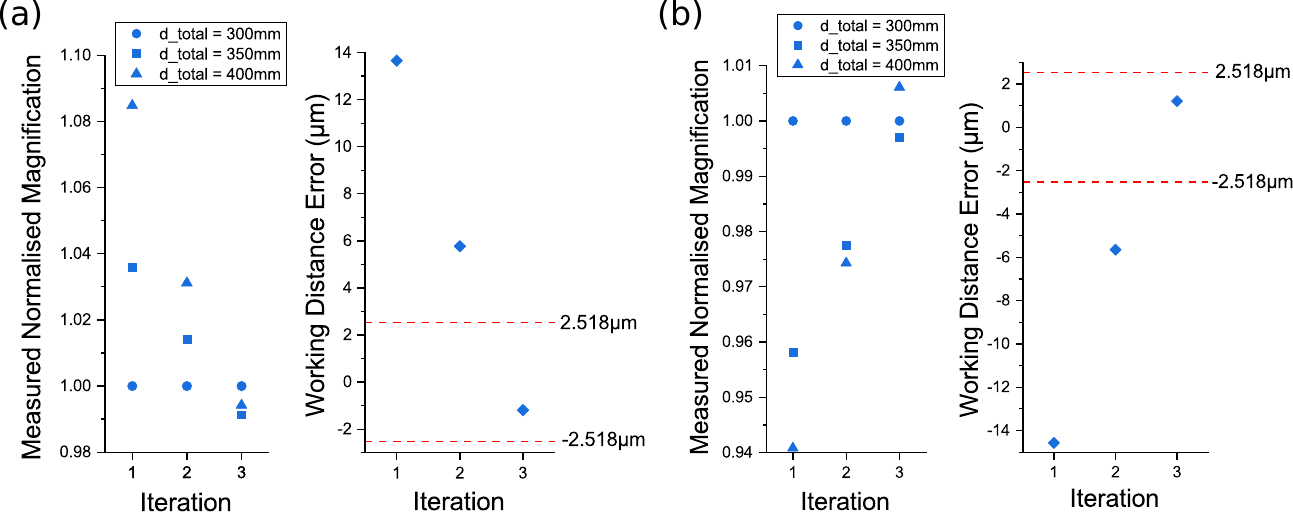}
    \caption{Calibration procedure results showing the measured normalised magnifications and the working distance errors for the situations where (a) the objective lens was intentionally positioned too far away from the sample and (b) too close to the sample. The red dashed lines on the working distance error graphs indicate the axial resolution limit of the M4All Fluorescence and TIE microscope, calculated using Abbe's axial resolution equation. Figure adapted from \cite{cairns_development_2022}.}
    \label{fig:calibrationinpractice}
\end{figure}

The alignment process was then iterated until the three channels had equal magnifications within the tolerances of the equipment and $d\_sample$ = $f_o$ within the optical axial resolution limit ($d_{z})$. For this example microscope the wavelength of light, $\lambda$, was 532\,nm, and the numerical aperture, $N\!A$, of the objective lens was 0.65, resulting in an axial resolution limit of 2.518\,$\mu$m using Abbe's axial diffraction equation, $d_{z} = 2\lambda / N\!A^{2}$. We carried out the alignment process for the situation where the initial position of the objective lens was intentionally too far away from the sample, and again when it was too close to the sample. The results are shown in figure \ref{fig:calibrationinpractice}. In both the cases shown it took three iterations of the alignment process to reduce the working distance error to less than the axial resolution limit (indicated by the red dashed lines on the working distance error graphs). Repeats for intentional misalignment of the objective lens and performing the alignment procedure can be found in supplemental figure 1. Please note, for transparency, the data in figure \ref{fig:calibrationinpractice} was obtained using an older version of our code where the RTMA model computations were performed using Maple\textsuperscript{TM} (Maple is a trademark of Waterloo Maple Inc.) and the magnification analysis was performed in a Jupyter Notebook, both of which are provided as supplemental material. We have since written both the RTMA model and analysis code in a single Jupyter Notebook which gives equivalent results and is also provided. The magnification plots in figures \ref{fig:SingleChannelplot} and \ref{fig:MultiChannelplot} were created using our new code.

\section{Conclusion}

Ray transfer matrix analysis within the paraxial approximation has been shown to effectively model the lateral magnifications of the imaging paths in a multichannel infinite-conjugate microscope when the optics are both aligned and misaligned along the optical axis. Furthermore, we have shown how magnification measurements from images acquired on each channel can be used to interpolate objective lens position from the model and how this information can be used to practically align the microscope optics. We have validated this alignment method on an open-source 3D printed multichannel microscope and shown it is a powerful tool when use of additional alignment hardware is not suitable (however, the method is applicable to all multichannel infinite-conjugate imaging systems). We provide the Python code for the ray transfer matrix analysis model and alignment algorithm as a detailed open-source Jupyter Notebook and believe it will be a useful tool for the open-source microscopy hardware community.

\subsubsection*{Data Accessibility {\normalfont All data and code underpinning this publication are available from Zenodo at\\ https://doi.org/10.5281/zenodo.8125287}}  

\subsubsection*{Competing Interests {\normalfont We declare we have no competing interests.}} 

\subsubsection*{Authors' Contributions {\normalfont G.S.C - conceptualization, data curation, formal analysis, investigation, methodology, software, validation and writing - original draft. B.R.P - conceptualization, funding acquisition, methodology, software, supervision, writing - original draft.}}  

\subsubsection*{Funding {\normalfont This work was funded under grants from the Royal Society (RGF\textbackslash EA\textbackslash 181058 and URF\textbackslash R\textbackslash 180017) and EPSRC (EP/M003701/1). When this work was carried out G.S.C. was funded under ‘OPTIMA: The EPSRC and MRC Centre for Doctoral Training in Optical Medical Imaging’ and B.R.P. held a Royal Society University Research Fellowship.}} 

\subsubsection*{Acknowledgements {\normalfont The work presented in this article originally formed part of Gemma S. Cairns' doctoral thesis at the University of Strathclyde \cite{cairns_development_2022}.}} 

\bibliographystyle{vancouver}
\bibliography{mainreferences}

\newpage
\section*{Supplemental Figure 1}

\begin{figure}[h!]
    \centering
        \includegraphics[width=15.25cm]{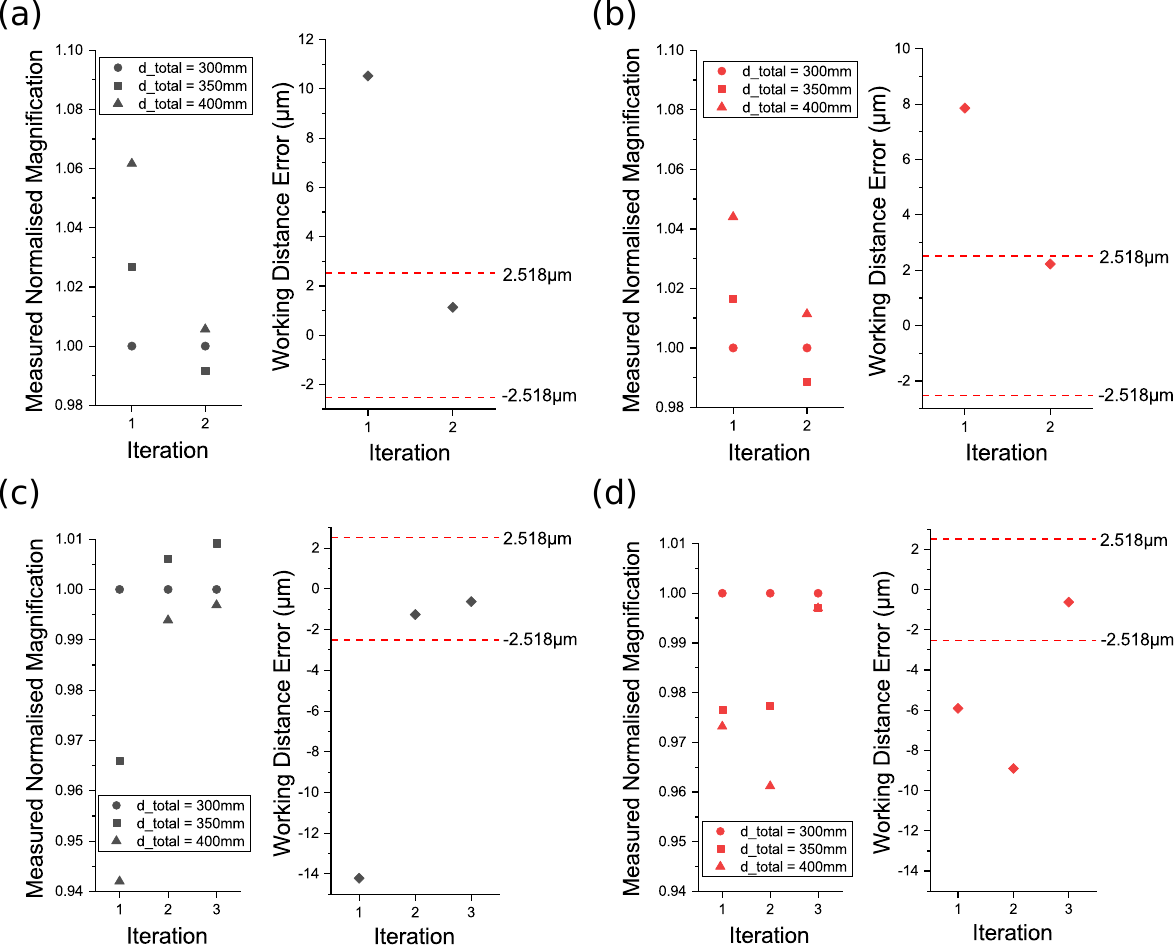}
    \caption*{Supplemental figure 1: Repetitions of the alignment routine carried out on the example M4All Fluorescence and TIE microscope for the cases where (a,b) the objective lens was intentionally positioned too far away from the sample, and (c,d) the objective lens was intentionally positioned too close to the sample. Figure adapted from [3].}
    \label{fig:supplemental1}
\end{figure}

\end{document}